How do the grains slide in fine-grained zirconia polycrystals at high temperature?

Maren Daraktchiev,[a)] and Robert Schaller

Ecole Polytechnique Fédérale de Lausanne, IPMC, CH-1015 Lausanne, Switzerland

Laurent Gremillard, Thierry Epicier, Jerome Chevalier, and Gilbert Fantozzi

Université de Lyon, INSA-Lyon, MATEIS - UMR CNRS 5510, F-69621 Villeurbanne, Cedex,

France

Degradation of mechanical properties of zirconia polycrystals is hardly discussed in terms of solution-precipitation grain-boundary sliding due to experimental controversies over imaging of intergranular amorphous phases at high and room temperatures. Here, the authors applied the techniques of mechanical spectroscopy and transmission electron microscopy (TEM) to shed light on the amorphization of grain interfaces at high temperature where the interface-reaction determines the behaviour of fine-grained zirconia polycrystals. They present mechanical spectroscopy results, which yield evidences of an intergranular amorphous phase in silica doped and high-purity zirconia at high temperature. Quenching of zirconia polycrystals reveals an intergranular amorphous phase on TEM images at room temperature.

---

[a)] Current address: Department of Earth Sciences, University of Cambridge, Cambridge CB2 3EQ, United Kingdom; Electronic address: mdar04@esc.cam.ac.uk

Mechanical loss or internal friction of a material is a quantity, which measures the conversion of the mechanical energy of a vibrating material into heat.[1,2] It is investigated by the technique of mechanical spectroscopy. In the framework of mechanical spectroscopy, the mechanical loss is examined by applying a periodic mechanical stress $\sigma(\omega)=\sigma_0\exp(i\omega t)$ and recording the dynamic response of the material $\varepsilon(\omega)= \varepsilon_0\exp(i\omega t-i\phi)$ under the effect of that stress, where $\varepsilon(\omega)$ is the strain as a function of circular frequency $\omega$. The phase lag $\phi$ or $\tan\phi$ (true for a small phase angle ϕ), which appears between the stress and strain is a measure of the internal-friction mechanism that dissipates the mechanical energy by motion of point defects, twin walls, grain-boundary interfaces, dislocations, relaxation into intergranular amorphous phases.[3-5] The evolution of crystal defects from one equilibrium state to a new one during the stress perturbation is ascribed to a peak appearance in $\tan\phi$ whose position and height account for the relaxation time $\tau$ and the relaxation strength of relaxing defects, respectively. From a microscopic point of view the relaxation strength is proportional to the concentration of crystal defects, while the relaxation time measures their mobility. As most of the dissipative mechanisms involve diffusion, then the peak in $\tan\phi$ can be employed in determining the activation enthalpy $Q$ and the attempt frequency $\tau^{-1}_0$ for a diffusion unit to overcome the local minimum of enthalpy in order to migrate towards another local minimum. A characteristic of thermal activation is the shift of peak position towards higher frequency with an increase of temperature during frequency measurements. $Q$ and $\tau^{-1}_0$ are calculated from an Arrhenius plot.[1,2] Alternatively, they can be also calculated from the Dorn method by constructing a master curve, which is a superimposition of $\tan\phi$ curves measured at different temperatures $T_i$ along the temperature compensated circular frequency axis, $\Omega=\omega.\exp(Q/R/T_i)$.[6] $Q$ is then derived from $\omega_1.\exp(Q/R/T_1)=\omega_2.\exp(Q/R/T_2)$, where $T_1$ and $T_2$ are two different temperatures. The attempt frequency is calculated from the position of the peak along the $\Omega$ axis. As a consequence,[1,2,6] the thermally-activated mechanical loss can be measured as a function of frequency by holding $T$ constant, as well as a function of temperature by holding $\omega$ constant.

The thermally-activated mechanical loss and dynamic modulus of high-purity 3 mol% yttria stabilized tetragonal zirconia polycrystals (3YS-TZP) as a function of temperature at a frequency of 1 Hz are reported in Fig.1. The high-temperature features of high-purity 3YS-TZP in Fig. 1 are the exponential increase in the mechanical loss, in parallel with the steep modulus decrease above 1300 K. By contrast, below 1300 K the loss spectrum is comparatively negligible as a function of temperature, in parallel with the saturation of modulus. Both spectra do not vary with temperature between 300 K and 1300 K, and for simplicity are not reported in Fig.1. The temperature of 1300 K is attributed to the onset of brittle-to-ductile transition[2] in high-purity 3YS-TZP. Above 1300 K the grains are submitted to minimal growth and elongation whereas as a whole the shape of grains remains unchanged.[7] Thereon, the transition from the state without dissipation (the brittle state) to the state with an exponential increase in dissipation (the ductile state) is mainly associated with grain-boundary sliding, as the fastest step, accommodated by climbing of grain boundary dislocations or solution-precipitation creep.[2,8] The accommodation mechanism prevents the grain-boundary steps, asperities, voids during the grain-boundary sliding and its nature originates from the microstructural features defined by the morphology and size of the grains, and the interface between the grains. In high-purity 3YS-TZP, in particular, the microstructure consists of grains with a 500 nm mean grain size; the grain interfaces are devoid of any amorphous phases at room temperature (see Fig. 2a). The lack of intergranular amorphous phase imposes the grain-boundary sliding controlled by climbing of grain boundary dislocations[8,9] as an underlying mechanism for the exponential increase in $\tan\phi$ in Fig.1. On the other hand, the grain boundary sliding controlled by solution-precipitation of matter in/from intergranular liquid phase[10-12] would be the most appropriate accommodation mechanism in the fine-grained polycrystalline ceramics having an amorphous film at the grain boundaries. Here, for the first time, we report mechanical-spectroscopy evidences of intergranular amorphous films in high-purity 3YS-TZP between 1100 K and 1600 K and discuss that discovery in the framework of mechanical-spectroscopy and high-resolution transmission electron microscopy (HRTEM) measurements of 3YS-TZP ceramics containing various amount of amorphous ($SiO_2$) phases. High-purity 3YS-TZP ceramics were processed by cold pressing and sintering from high-purity 3Y-TZP powders. The post-sintered samples have a density of 97% of theoretical density. $SiO_2$ doped 3YS-TZP samples were prepared by slip-casting method.[13] The spectroscopy measurements were carried out in an inverted torsion pendulum, working with forced vibrations.

It has been largely communicated in literature that ceramics processed by either hot or cold pressing from powder compacts with sintering aids manifest high mechanical loss and low elastic modulus attributed to dissipation in the intergranular amorphous phases.[5,14-17] The sintering aids added to the powder compacts for improving the powder compressibility and densification during the processing, together with powder impurity, result in the formation of amorphous phases at triple junctions and grain boundaries. Indeed, a few percent of sintering aids in $Si_3N_4$ ceramics, in particular, instigates large amorphous-phase agglomerates (amorphous pockets) at multiple junctions and thin or even thick amorphous film at grain boundaries.[14,16,17] By contrast, doping of 3Y-TZP ceramics with $SiO_2$ phases does not manifest any amorphous films at grain boundaries. This is attested in Fig. 2b by HRTEM images showing a perfect coincidence of crystallographic planes at grain boundaries in $SiO_2$ doped 3SY-TZP. Amorphous phases are only found to accumulate as amorphous pockets at triple points in $SiO_2$ doped 3Y-TZP (Fig. 2c); a rigid skeleton of grains surrounds the pockets. Moreover, extensive research has shown that the larger the amount of $SiO_2$ aids, the larger the pocket volumes and the rounder the grain edges in contact with the pockets.[13,18]

The mechanical-spectroscopy spectrum of $SiO_2$ doped 3YS-TZP is a fingerprint of high-temperature dynamics of the amorphous pockets and the grain skeleton surrounding the pockets. Therefore, two spectroscopy features in 3YS-TZP with 2 wt% of $SiO_2$ phases are seen in Fig. 1. They are the relaxation peak at 1410 K and the exponential increase in energy dissipation (exponential background) above 1460 K, respectively. Both the peak and the background are accompanied by a steep decrease of shear modulus. More specifically, the peak is a manifestation of dissipation in the amorphous phases above glass transition temperature. It shifts towards high temperature with a decrease of $SiO_2$ content and disappears in high-purity 3YS-TZP, which does not contain any $SiO_2$ aids. The peak in Fig. 1 gives for $Q$ and $\tau^{-1}_0$ values of 640 kJ/mol and $10^{24}$ s$^{-1}$, respectively. The exponential background is a manifestation of the deformation of the rigid skeleton around the pockets.[17] It dominates over the peaks in $SiO_2$ doped 3YS-TZP at high temperature, as well as $\tan\phi$ of high-purity 3YS-TZP above 1300 K at 1 Hz, where the mechanical loss occurs by deformation of grains in the skeleton. During mechanical spectroscopy at 1 Hz frequency range the experimental conditions are close to transient creep, and therefore only local motions of grains in the skeleton can take place, the easiest ones are the relative displacements of neighboring grains.[12,19] Furthermore, the sliding of the grains in the skeleton will be controlled by the opposing stresses built up at grain-multiple junctions caused by the elastic deformation of opposing grains at triple junctions. It will also be controlled by the property of interfaces between the grains, where a few-nanometer thick liquid film attributed to impurity, as well as sintering aids, has been predicted by Clarke[20] in 1987. Clarke[20] has calculated its equilibrium thickness from the attractive Van der Waals forces between the grains on both sides of the intergranular film and the repulsive force due to the microstructure of the liquid film at high temperature.

In the process of grain-boundary sliding at high temperature, the intergranular film in ceramics will play an important lubricating role. On the one hand, the grain boundary sliding accommodated by diffusion through that film would allow the enhanced ductility or even structural superplasticity of high-purity fine-grained ceramics without any assistance of dislocations. Dislocation activity is seldom observed inside the grains in those ceramics. On the other hand, the intergranular film lubricating the grain-boundary sliding, enhancing the deformation and shifting the brittle-to-ductile transition in ceramics towards lower temperature is recognized as a main mechanism for failure of mechanical properties of fine-grained ceramics at high temperature. Very small amounts of amorphous additives are able to modify the grain interfaces and the grain morphology, and even change dramatically the mechanical behavior of ceramics. Recently Hiraga *at al*.[21] have shown that an amount of 0.3 wt% of $SiO_2$ additives into the 3Y-TZP powder result in a significant increase of plastic deformation at high temperature.

As no intergranular amorphous films are imaged at room temperature,[13,18,21] we shall explore the idea that the liquid film at high temperature[20] might be conserved around the grains at low temperature by applying high rates of cooling. During such a non-equilibrium cooling process the local thermodynamic equilibrium conditions will be violated in 3YS-TZP ceramics; hence the intergranular liquid film may vitrify without crystallization at grain boundaries or rejection at triple junctions. Therefore, one would expect that the higher the cooling rate, the higher the probability of observing that film at room temperature. The intergranular amorphous phase at room temperature would be then in a close relationship with the conditions of cooling in 3YS-TZP. The respective co-workers[13,22] have recently shown that depending on the post-sintering cooling rates in $SiO_2$ doped 3YS-TZP ceramics two phenomena can occur at grain interfaces. The grain interfaces are either free of any intergranular phases (Fig. 2b) when the cooling rate is relatively low[13] or wetted by intergranular amorphous films (see Fig. 2d) when quenching is carried out.[22] The amorphous film in Fig. 2d is rich of silica with a thickness of 1.8 nm, which is a value close to the equilibrium thickness' predicted by Clarke.[20]

The comparison between as-sintered and quenched $SiO_2$ doped 3YS-TZP is shown in Fig.3. The mechanical spectra reveal an exponential increase of $\tan\phi$ above 1100 K in both samples. Note that we plot the experimental data against the logarithmic scale in Fig. 3 giving a linear dependence of $\tan\phi$ as a function of temperature. No difference in $\tan\phi$ is seen in as-sintered and quenched $SiO_2$ doped 3Y-TZP below 1100 K and above 1380 K. Below 1100 K, the energy dissipation is negligible because the grain-boundary sliding is not activated by temperature; above 1380 K, however, the grain boundary sliding is activated by temperature and both spectra overlap. Spectral overlapping adumbrates that the microstructures of quenched and as-sintered $SiO_2$ doped 3YS-TZP evolve towards the same thermodynamic state at high temperature. Between 1100 K and 1380 K, both microstructures are different; hence, both spectra are different. Further, $\tan\phi$ reveals a relaxation peak ($Q = 600(\pm 40)$ kJ/mol and $\Omega = 10^{23(\pm 1)}$ s$^{-1}$) at 1225 K in quenched samples. Building on the HRTEM images in Fig. 2d and the value of activation energy, being close to grain-boundary diffusion for $Zr^{+4}$-ions and $Y^{+3}$-ions[7,23] in 3Y-TZP, we can associate the peak in the quenched samples with a dissipation mechanism operating between the intergranular film and grain interfaces.[10]

Insights into the relaxation mechanisms in high-purity 3YS-TZP can also be found in Fig. 3. Similarly to SiO$_2$ doped 3Y-TZP, we compare the spectrum of high-purity 3YS-TZP without any quenching (as-sintered) with the one measured after post-sintering quenching (quenched) in order to quantify the changes in the mechanical loss spectrum due to modification of the grain interfaces during quenching. Again, no difference in tanϕ has been observed below 1200 K and above 1500 K. Below 1200 K, both spectra are comparatively small; above 1500 K, they are identical because the microstructures are also identical. Between those temperatures, however, both spectra are different. The quenched sample shows higher energy dissipation than the as-sintered one with a relaxation peak at about 1345 K at 1 Hz. The activation parameters were found to be similar to those measured in the quenched SiO$_2$ doped 3YS-TZP. Note that the amount of SiO$_2$ additives is less that 0.004 wt% in high-purity 3YS-TZP powders (Tosoh Corporation, Japan). This amount seems to be too small for a film formation at high temperature, but silica is not the only one compound that is responsible for the amorphous interfaces. Other impurities, such as Al$_2$O$_3$ (<0.005wt%) and Na$_2$O (0.005 wt%), exist in 3YS-TZP as well and some of them are known to form liquid phases with SiO$_2$ at high temperature (Yttrium and Zirconium[22] may also be found in glassy phase). Those phases may transform into glassy ones during quenching, and therefore the peak at 1345 K may be attributed to the residual amount of the intergranular phases after quenching. The mechanical loss measurements in high-purity 3YS-TZP, as well as the mechanical loss measurements and HRTEM observations in quenched SiO$_2$ doped 3YS-TZP, support the conclusion of solution-precipitation[10] controlled grain-boundary sliding in 3YS-TZP ceramics at high temperature.

In Fig.3, the spectra of high-purity 3YS-TZP are shifted towards higher temperatures with respect to SiO$_2$ doped 3YS-TZP. The reason is in the lubricating efficiency of the SiO$_2$ glassy phase and in the different mean grain sizes $d$ that exists between high-purity and SiO$_2$ doped 3YS-TZP. The former has $d \approx 0.58$ μm, the latter $d \approx 0.38$ μm. The larger grain sizes manifest lower energy dissipation by grain-boundary sliding.[2,12,24] The grain sliding kinetics in the high-purity ceramics will be slowed down by larger grain size and lower content of lubricating agent at the boundaries. The shift of the peak towards higher temperature in high-purity 3YS-TZP evinces the amount of energy in the form of heat that must be supplied to $SiO_2$ doped 3YS-TZP to obtain identical displacement of zirconia grains in both polycrystalline ceramics.

In conclusion, the high temperature mechanical property of 3YS-TZP ceramics is determined by the grain size, grain morphology and local thermodynamic states, which favor the formation of liquid intergranular films. Decreasing of temperature supports either film absorption by grains or film rejection at multiple grain junctions. Thereon, no film will be imaged at grain boundaries in zirconia ceramics at low temperature unless quenching from high temperature is achieved.

This work was supported by the Swiss National Science Foundation.

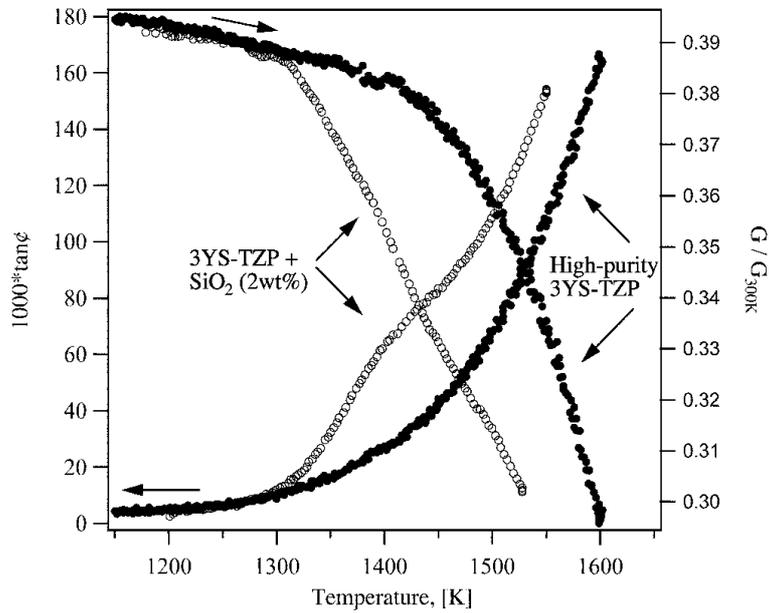

FIG 1. Mechanical loss tanϕ and dynamic shear modulus $G/G_{300K}$ as a function of temperature for two grades of 3YS-TZP. An exponential increase of high purity 3YS-TZP (grain size 0.58 μm) above 1300 K; a relaxation peak of $SiO_2$ doped 3YS-TZP (grain size 0.3 μm) at 1410 K. Steep decrease of moduli in both samples.

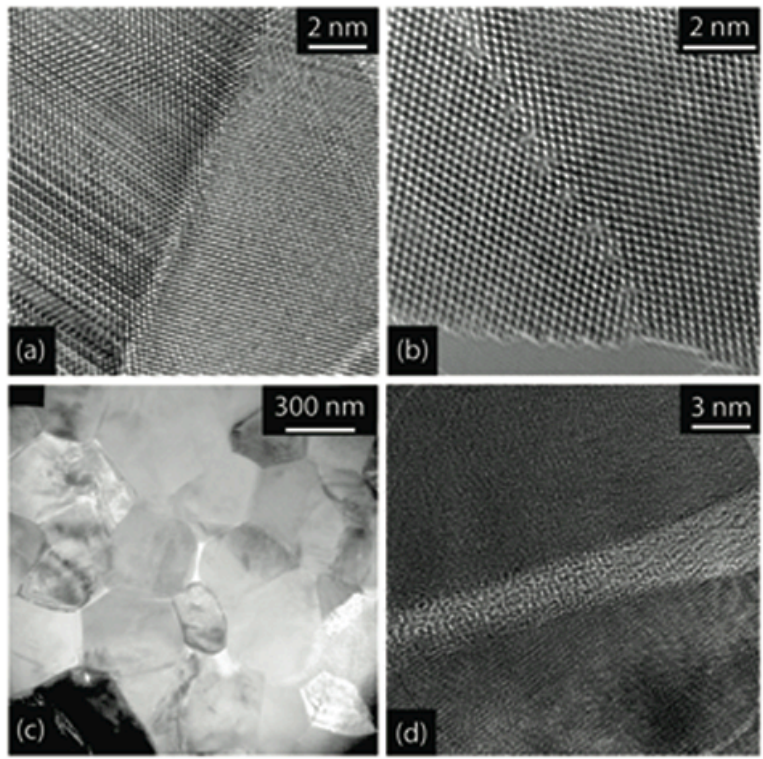

FIG 2. HRTEM micrographs of (a) grain boundary in high-purity 3YS-TZP, (b) grain boundary in SiO$_2$ doped 3YS-TZP; (c) dark field conventional TEM of large SiO$_2$ pockets in SiO$_2$ doped 3YS-TZP, and (c) HRTEM image of intergranular amorphous film in SiO$_2$ doped 3YS-TZP after quenching.

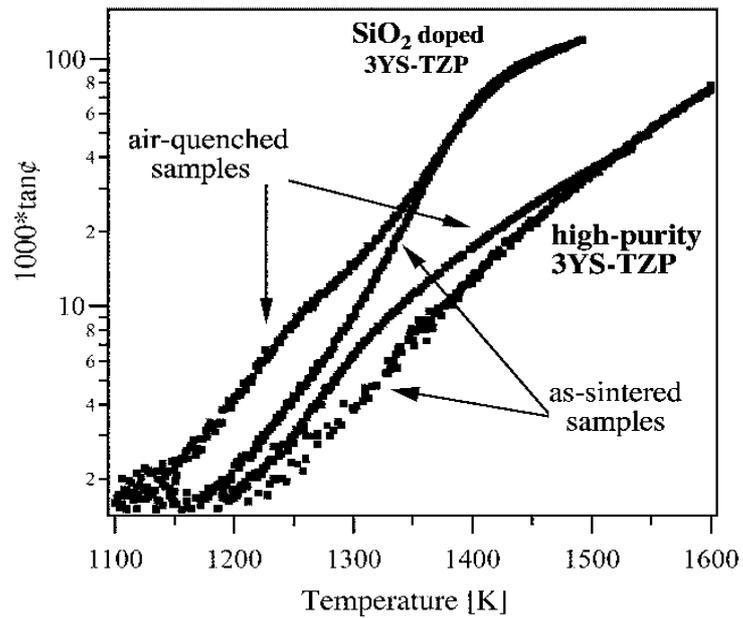

FIG 3. tan$\phi$ of high-purity and SiO$_2$ doped 3YS-TZP. tan$\phi$ increases after quenching in both 3YS-TZP ceramics. The temperature rate and frequency are 1 K/min and 1 Hz, respectively. The shift of high-purity 3YS-TZP towards higher temperatures is attributed to grain size and grain morphology.